\documentclass[11pt]{article}
\usepackage{fa2023}
\usepackage{amsmath}
\usepackage{cite}
\usepackage{url}
\usepackage{graphicx}
\usepackage{color}
\usepackage{siunitx}
\usepackage[utf8]{inputenc}
\usepackage{subcaption}
\usepackage{float}

\title{3D Audio-Visual Recordings of Mosquito Wings for Aeroacoustic Simulation}

\multauthor
{Lionel Feugère$^{1,2*}$ \hspace{1cm} Jung-Hee Seo$^3$ \hspace{1cm} Umair Ismail$^3$} { \bfseries{Gabriella Gibson$^1$ \hspace{1cm} Rajat Mittal$^3$ \hspace{1cm}}\\
  $^1$ Natural Resources Institute University of Greenwich Chatham, Kent ME4 4TB, UK\\
$^2$  L2TI, Université Sorbonne Paris Nord, F-93430, Villetaneuse, France \\
$^3$ Department of Mechanical Engineering Johns Hopkins University Baltimore, MD, 21218, USA\\
\thanks{ *\textit{\textbf{Corresponding author}: lionel.feugere@riseup.net. \endgraf \noindent \textbf{Copyright}: \textcopyright\the\year{} L. Feugère et al. This is an open-access article distributed under the terms of the Creative Commons Attribution 3.0 Unported License, which permits unrestricted use, distribution, and reproduction in any medium, provided the original author and source are credited.}}
}

\begin{document}

\maketitle
\begin{abstract}
Mosquito acoustic communication is studied for its singular and poorly-known in-flight hearing mechanism, for its efficiency in mechanical-to-acoustical power transduction, as well as for being the deadliest disease vector. A combined computational and experimental methods to predict and extract the wing-tone sound from individual tethered or free-flying mosquitoes was developed. This paper describes the experimental methods and gives some preliminary results of the simulations. Simultaneous 3D video and 3D sound of \textit{Culex quinquefasciatus} mosquitoes were recorded. The sound map around the mosquitoes was recorded in one or two planes with a rotating array of 12 microphones. Back-illuminated mosquito-wings allowed to extract 11 vein-crossing locations on each camera image (20,000 frames per second) over 3‐4 wingbeat periods to generate 3D deformations of the wing. Sound data recorded by microphone arrays were post‐processed by using the physics-based independent component analysis to filter out the noise and generate a 3D sound map. The simulated wing-tone sound pattern generated from the aeroacoustic simulation agrees well with the original recording in the experiment using the microphone array. The methods we developed will allow us to investigate the wing-tone soundscape of individual mosquitoes during the courtship and mate-chasing.
\end{abstract}

\section{Introduction}\label{sec:introduction}

The aerial courtship dance of mosquitoes has fascinated entomologists for more than two centuries \cite{Deriville60,Knab06}. The chase between female and male in mating swarms involves highly controlled variations in the frequency and intensity of flight‐tones (i.e. sounds generated by the flapping wings) \cite{GibsonRussell06, WarrenGibRus09,SimoesIngGibRus16,PantojasanchezGVAA19, SomersGSBAAMNMSA22} with concurrent changes in flight speed and direction \cite{PuckettRuiOuellette15,NakataSimWalRusBom22}, and enables detection of conspecifics\cite{FeugereRouxGibson22,NakataSimWalRusBom22}, possible display of fitness and transmission of mating interest \cite{CatorNghHoyHar10,AldersleyCator19}. However, despite over a century and a half of research \cite{FeugereSimRusGib22}, significant knowledge gaps continue to exist in our understanding of this behavior. To decipher this courtship dance, entomologists must integrate acoustic, energetic and flight information for untethered, free‐flying mosquitoes, but the tools that can provide these data have, so far, been limited \cite{BomphreyNakPhiWal17,NakataSimWalRusBom22}. In the present study, we combine the computational biomechanics and acoustics, and behavioral entomology, to generate data and insights into the biomechanics and physics of courtship‐associated acoustic communication in mosquitoes. By combining computational modeling with biological assays, we generate frequency polar pattern of mosquitoes engaged in mating-swarm behaviour. 
%

\begin{figure}[H]
 \centerline{\framebox{
 \includegraphics[width=1\columnwidth]{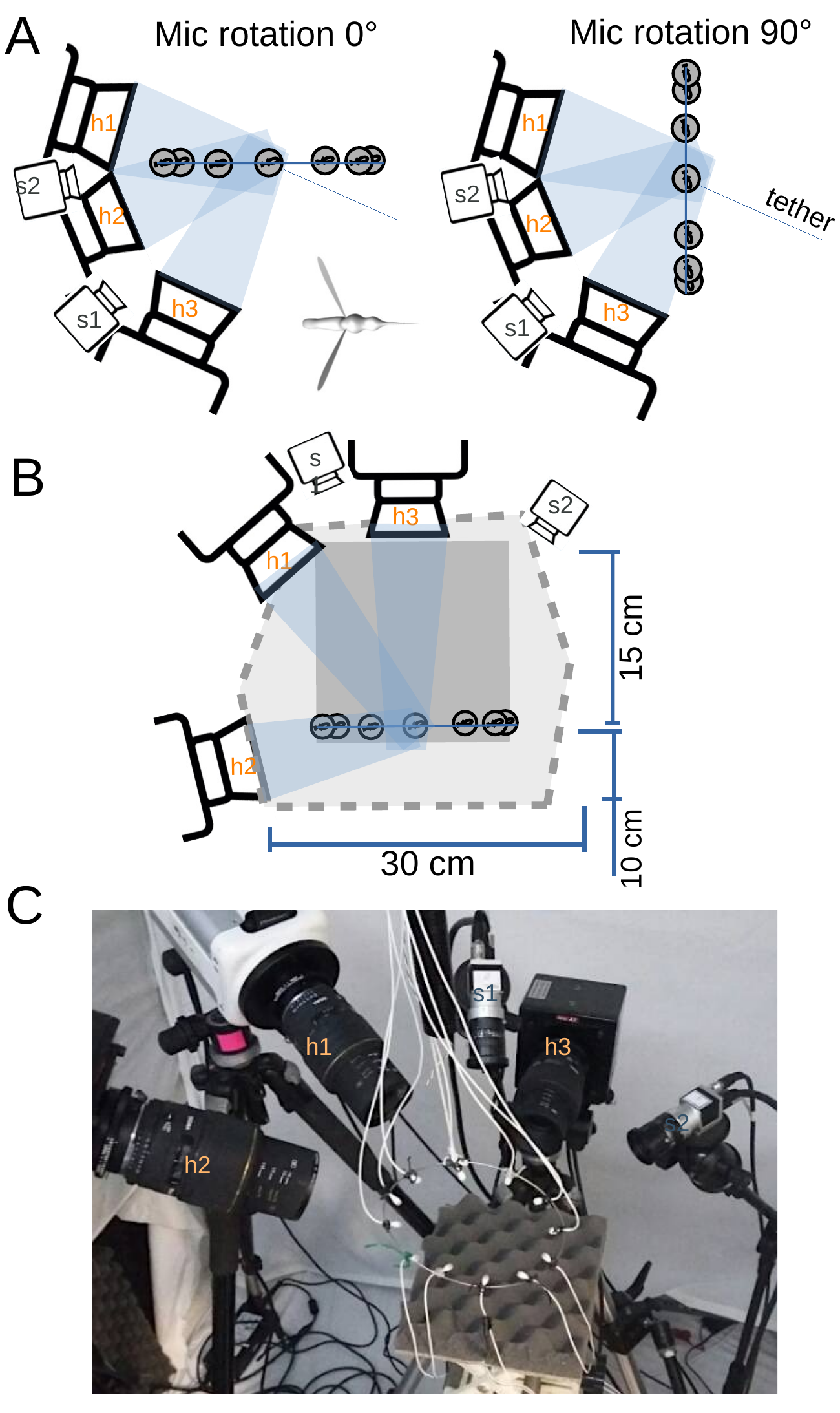}}} 
 \caption{Experimental set-up for tethered (A) and free-flying mosquitoes (B,C). High-speed cameras (h1, h2, h3) are directed toward the centre of a microphone array back-lighted by IR-lights, while still-cameras (s1, s2) record the space around the whole circular microphone array. A: Schematic top view of the tethered-mosquito setup with a rotating microphone array with fields of view of each high-speed camera in blue. B: Schematic top view of the free-flying setup, with fields of view of each high-speed camera in blue, net boundaries in dashed lines and square swarming marker in grey. C: Free-flying mosquito setup (without the net) with an acoustic foam used as a swarming mosquito marker.}
 \label{fig:setup}
\end{figure}

\section{Method}

\subsection{Mosquitoes}

All experiments were performed with virgin \textit{Culex quinquefasciatus} Say, 'Muheza' strain (separated by sex $<$ 24h post-emergence). The colony was established at the Natural Resources Institute, University of Greenwich (UK) from eggs provided by the London School of Hygiene and Tropical Medicine, UK ,and kept in environmentally controlled laboratory rooms with a 12h:12h light:dark cycle, $>$~60\% relative humidity and $\sim$ 24--26~$^{\circ}$C. 

The experiments were performed in July 2020 with tethered mosquitoes and in October 2021 for free-flying mosquitoes, at a time of the day corresponding to their circadium swarming phase (-~1~h to +~3~h after the beginning of the 12~h dark cycle of the rearing room for the tethered mosquitoes and between 0.3~h and 1.3~h for the free-flying mosquitoes), in the presence of a light setup mimicking the sunset as in~\cite{FeugereRouxGibson22}. For the tethered setup, a total of 13 males and 13 females were selected to be recorded based on their ability to beat their wings without interruption. Mosquitoes were tethered by warming bee wax on the tip of a thin wire which was kept right after on the top of their abdomen until the wax hardened. The other tip of the thin wire was attached to a syringe needle, which was fixed to a boom with adhesive putty. This let them beat their wing freely while being immobile and let the experimenter adjust their position. They were recorded in an individual configuration (i.e., no other mosquito around). 

For the free-flying setup, 2- to 11-day-old males and females were released in the recording cage. Only individual males and females flying alone were selected for the analysis of this paper.  A 15 cm $\times$ 15 cm grey acoustic foam was used as a visual marker for the free-flying mosquitoes to swarm around the audio-video recording spot (\figref{fig:setup}). In \textit{Culex quinquefasciatus}, the swarming behaviour of a single or several mosquitoes consists of a loop trajectories next to a visual marker~\cite{Gibson85}. When adding the microphone array above the rectangular acoustic foam, the swarm centre moved to the foam side which pointed toward the room lights, whatever the microphone array position, so the array was moved on this side to have the swarm centre at the centre of the array.

\subsection{Audio/video recordings}

\textbf{High-speed cameras. }Three high-speed cameras (FASTCAM Mini AX200, Mini AX100 and Nova S6
) were fitted with 105 mm macro lens (Zoom, Navitar) and an IR LED light (HAY-IR-70/30 or Bosch UFLED20-8BD) was located in front of each camera, with the mosquito at a distance of 20--22 cm from the camera lenses for the tethered setup  and 15--18 cm for the free-flying setup (\figref{fig:setup}). Video resolutions were $384\times384$ pixels with 20k images per second for tethered mosquitoes and $512\times512$ pixels with 22.5k images per second for free-flying mosquitoes. For tethered mosquitoes, the focal points were set on the mosquito’s left wing. The recording was triggered manually and images were recorded under the software provided by the camera (Photron FASTCAM Viewer v4).

\textbf{Microphones. }For the wing-tone recording, twelve omnidirectional miniature microphones (DPA 4060; frequency range of 0.02--20 kHz with a 3 dB boost at 8--20 kHz) were spread over a 9cm-radius circular array (30$^{\circ}$ step) with a metal frame hung at the ceiling (\figref{fig:setup}) and plugged to a sound interface (MOTU Stage B16 runned on MacOSX 10.13.6). Sounds were recorded on a digital audio workstation (AudioDesk 4.01 for the tethered setup, Performer Lite 10.13 for the free-flying setup) at 24 bits, and sampling rate of 192 kHz for the tethered mosquitoes and 48 kHz for the free-flying mosquitoes. The cameras were located as far as possible from the preparation to limit sound reflections on the lens: the lenses were at a minimum distance of 11 cm from the closest microphone depending on the microphone and the array rotation (\figref{fig:setup}) for the tethered setup and a minimum distance of 6 cm for the free-flying setup (\figref{fig:setup}B, C). For the tethered setup, the microphone array was rotatable around the vertical axis. Four rotations were performed for each mosquito (0$^{\circ}$, 30$^{\circ}$, 60$^{\circ}$ and 90$^{\circ}$) but only two (0$^{\circ}$, 90$^{\circ}$; \figref{fig:setup}A) were post-processed and included in the results. 

\textbf{Audiovisual synchronization. }High-speed camera and sound recordings were synchronized with the camera's synch pulse sent from one of the camera to the sound interface. Synchronization sampling error was limited by the high-speed camera sampling period, i.e., 50~$\mu$s. Due to the speed difference between sound and light, at a given time, the recorded sound corresponded to the sound generated by the wing at a distance of 9 cm, i.e. 265~$\mu$s before.

\textbf{Large-field-of-view cameras and calibration. }Two cameras (Basler model-acA640-120gm for the tethered setup; acA2040-90umNIR for the fre-flying setup) were fitted with wide-angle lenses (Computar T3Z3510CS for the tethered setup, Fujinon HF6XA-5M for the free-flying setup) to capture a picture of the whole microphone-array from two different views. They were plugged to a personal computer (running on Windows 10) and recorded (Pylon Viewer 6.2). Camera calibration were performed simultaneously for high-speed cameras and wide-angle cameras by the help of a set of \textregistered Lego pieces, for each recording. 

\textbf{Temperature. }Room temperature ranged between $25-28^\circ$C, measured with a logger (HH506RA, Omega Engineering, Inc) and a thermocouple Type T, IEC 584 Class 1 (Omega Engineering, Inc) just above the microphone array during each recording, with  a total measurement accuracy error of $\pm 0.9^\circ$C.

\begin{figure}[!h]
   \begin{subfigure}[b]{1\columnwidth}
 \centerline{\framebox{
 \includegraphics[width=1\textwidth]{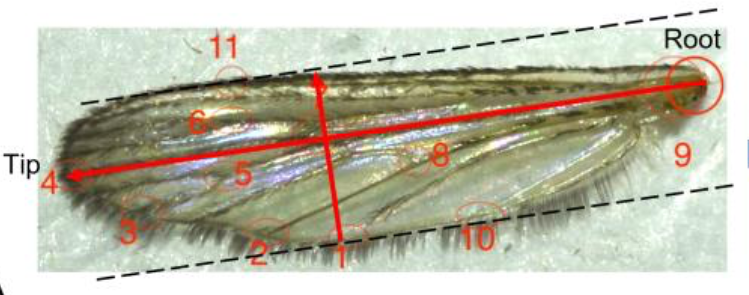}}}
 \caption{}
    \end{subfigure}\\
   \begin{subfigure}[b]{0.65\columnwidth}
 \centerline{\framebox{
 \includegraphics[width=1\textwidth]{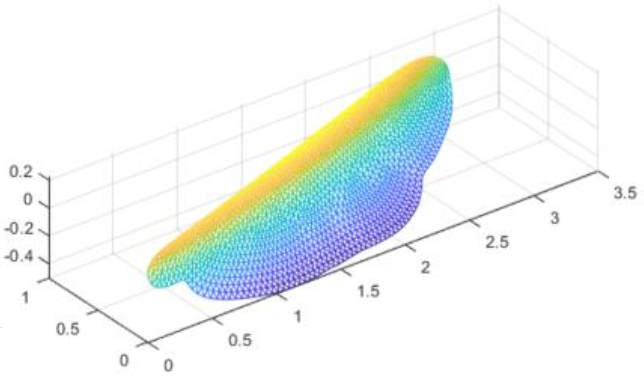}}}
 \caption{}
    \end{subfigure}~~~~
       \begin{subfigure}[b]{0.32\columnwidth}
 \centerline{\framebox{
 \includegraphics[width=1\textwidth]{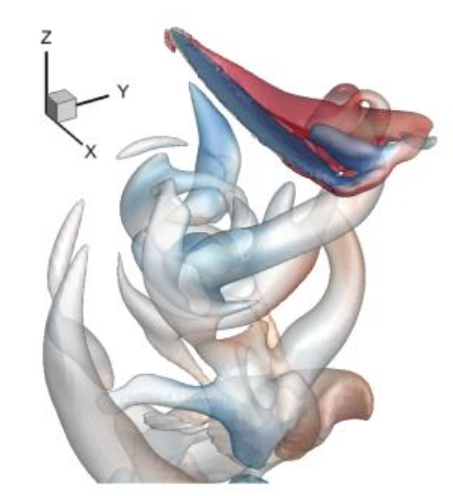}}}
 \caption{}
    \end{subfigure}\\
       \begin{subfigure}[b]{1\columnwidth}
 \centerline{\framebox{
 \includegraphics[width=0.7\textwidth]{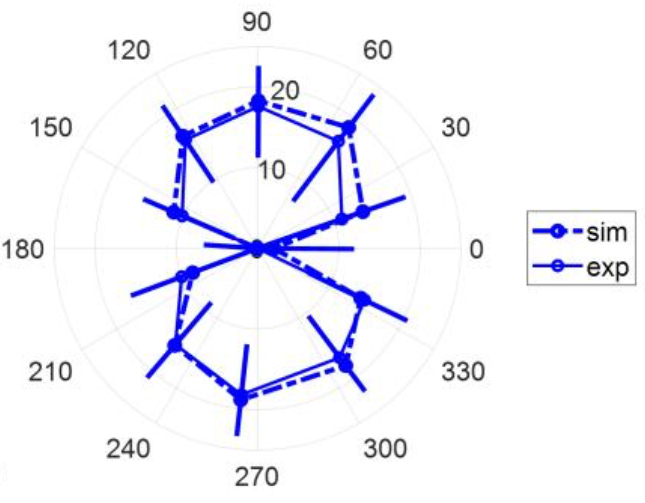}}}
 \caption{}
    \end{subfigure}\\
 \caption{$(a)$ Mosquito wing and 11 landmark points. $(b)$ Reconstructed 3D wing model. $(c)$ Aerodynamic simulation results for a representative tethered mosquito case. Instantaneous vortical structure visualized by the iso-surface of Q. $(d)$  $f_0$ sound directivity pattern for a tethered specimen. Dashed line: simulation result; solid line: experimental measurements by the microphone array; measured sound pressure is processed by using the Green’s function based ICA method.}
 \label{fig:sim}
\end{figure}

\subsection{Wing kinematics reconstruction}

To reconstruct the 4D wing kinematics, which is required for the simulation, 11 landmark points on the mosquito wings were tracked from the high-speed camera recording data (\figref{fig:sim}a). 
For each mosquito recording, these points were semi-manually recorded using the DLTdv software \cite{Hedrick08} over $\sim$200 images ($\sim$ 3--4 wingbeat periods) for each of the three high-speed cameras.
The wing kinematics and deformations were extracted by applying the least-square-fitting surface method. In addition to the wing twisting, span wise and chord wise bending are also considered for the wing deformation. The out-of-plane wing deformation is fit into a second-order polynomial surface (\figref{fig:sim}b).

%

\subsection{Physics based independent component analysis}
The 3D sound data recorded from the experiment are processed by applying a novel physics-based independent component analysis. The method separates out the physically generated sound by using the free-space Green's function and thus is very effective to filter out background noise. Assuming the wing-tone sound pressure recorded by each microphone, $p'_j$, is generated by the physical, dipole source, \textbf{F}, and is contaminated by the statistically same background noise, $\epsilon$, the physical dipole source and noise component in the frequency domain can be found by using the singular value decomposition (SVD) of the Green’s function matrix:

\begin{align}
	\begin{bmatrix} 
		 \textbf{F}(\omega)\\
		 \epsilon(\omega)  
	\end{bmatrix}	 
	 &  = \left[ \textbf{G}_j(\omega) ~~~~1 \right]^+ \left[ p_j '(\omega) \right]
 \end{align}

where $\textbf{G}_j$ is the free-space Green's function from the source (mosquito) location to the $j$-th microphone position, and $+$ denotes a pseudo-inverse of the matrix. The physical wing-tone, $p'_{w,j}$ can be then obtained by $[p'_{\omega,j} (\omega)]=[\textbf{G}_j(\omega)][\textbf{F}(\omega)]$.
The processed 3D sound data showed a typical dipole sound pattern which is a well-known characteristics of wing tone sounds.

\subsection{Aerodynamic and aeroacoustic simulation}

The CFD simulations presented herein are performed for single free-flying mosquitoes using wing kinematics reconstructed from the experimental recordings. These computations are performed in a reference frame affixed on the body of a translating mosquito. The mosquito wing, which is discretely represented via triangular mesh elements, is modelled by a zero-thickness membrane using a sharp-interface immersed-boundary method ~\cite{MittalDBNV08} and its motion is prescribed using the reconstructed wing motion data. Further details on the reconstruction of the wing motion will be discussed elsewhere.

\begin{figure}[H]
    \centering
    \begin{subfigure}[b]{1\columnwidth}
 \centerline{\framebox{
 \includegraphics[width=0.7\textwidth]{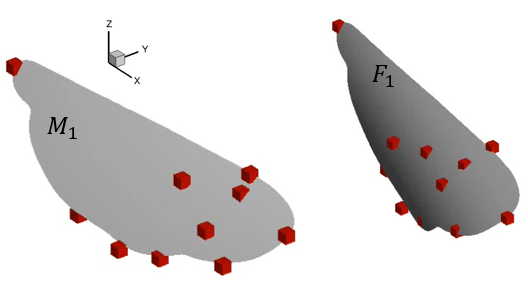}}}
 \caption{}
    \end{subfigure}\\
    \begin{subfigure}[b]{1\columnwidth}
 \centerline{\framebox{
 \includegraphics[width=0.7\textwidth]{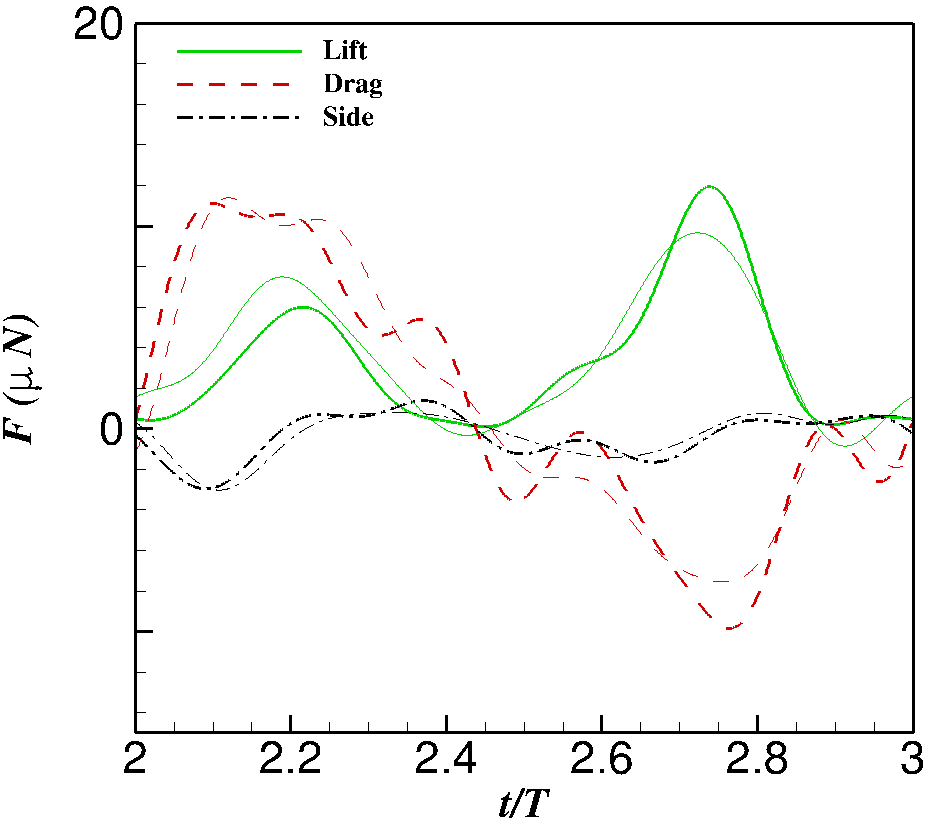}}}
 \caption{}
    \end{subfigure}\\
    \begin{subfigure}[b]{1\columnwidth}
\centerline{\framebox{
 \includegraphics[width=0.7\textwidth]{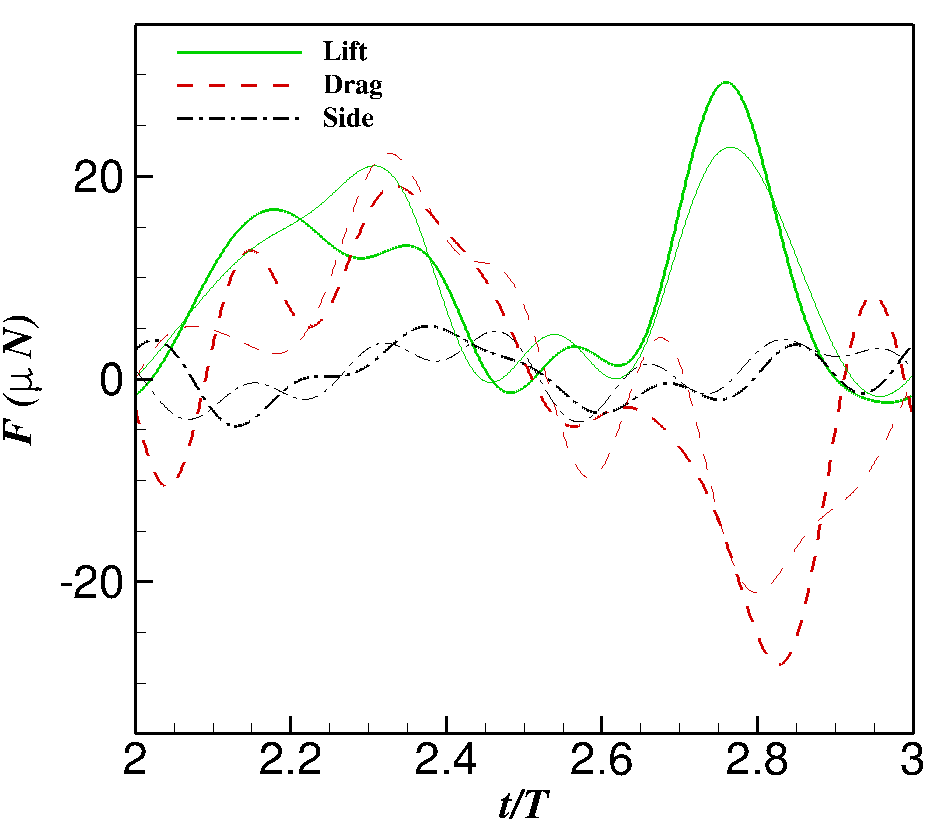}}}
 \caption{}
    \end{subfigure}
   \caption{$(a)$ An instantaneous snapshot of the wing during the flapping cycle upon construction from experimental data and for use in CFD simulations. The position of the 11 points tracked during the experiment at the  same instant is also included using red cubes. Temporal variation of the pressure-induced lift, drag and side force along the flapping cycle for $(b)$ $M_1$ and $(c)$ $F_1$. Thick and thin lines in $(b,c)$ represent the results from two distinctly constructed wing kinematics that correspond to a different cycle in the recording flapping motion. Downstroke: $2.0 \le t/T \le 2.5$; upstroke: $2.5 \le t/T \le 3.0$.}
    \label{fig:force}
\end{figure}

The anterior end of the mosquito is pointed in the $-y$ direction. As a result, its nearly horizontal flapping stroke lies in the $x-y$ plane (\figref{fig:force}a.) and the direction of the vertical weight-supporting force is $+z$. The wing is immersed into a Cartesian volume grid of size $L_x \times L_y \times L_z = 5L_s \times 10L_s \times 10L_s$ ($L_s$ is the spanwise length of the wing). The flow domain is divided into $N_x \times N_y \times N_z = 193 \times 257 \times 257$ grid points. A uniform albeit fine mesh, with a grid size of $0.01L_s$, is employed in the section that covers the flapping volume and the region penetrated by flapping-induced vortices. Neumann boundary conditions for all three components of the velocity vector are applied at five out of the six faces of the flow domain. As the aerodynamic study investigates the flow physics and the vortical patterns induced by one flapping wing, a symmetry condition is instead imposed on the sagittal plane ($x=0$). The computational time step is fixed at $\Delta t=0.001/f_0$, which resolves one wing-beat cycle in 1000 time steps and ensures that the Courant–Friedrichs–Lewy condition is below $0.5$. All CFD simulations are performed for five wing-beat cycles and only the results from the final cycle are used for analysis. However, it is important to note that the computed force/power levels and the examined coherent structures are virtually identical in cycles 2-5.

The aeroacoustic sound generated by the wing's flapping motion is predicted by employing the well-established formulation for the moving surface – the Ffowcs Williams and Hawkings (FW-H) equation. The FW-H equation has been widely used for the prediction of aeroacoustic sound from rotating blades, and also for the sound from flapping wings. We employ an integral formulation of the FW-H equation  \cite{KennethFarassat97}. The surface pressure distributions on the wing obtained from the aerodynamic simulation are recorded at every time step, and the sound pressure is calculated by the FW-H equation from the surface pressure and velocity data. 
More details about the simulation methods can be found in \cite{SeoHedrickMittal19}.

\section{Results}

To verify and validate the reconstruction, data processing, and simulation methods, the described methods are first applied to a tethered mosquito. The wing kinematics and deformation are reconstructed from the high-speed video recording data, and the aerodynamic simulation is performed. Within the tethered subclass, the wing-tone sounds at the micro-phone array locations as predicted by the FW-H equation are compared against the measured ones to validate the methods.

The recorded wing-tone sound, however, did not show a typical dipole pattern because of the background noise. At the distance of 9 cm, the wing-tone sound was as weak as about 20 dB at the peak and thus the signal-noise ratio was low. To extract the physical wing-tone sound, the physics based independent component analysis (ICA), and the processed sound data showed a typical dipole pattern which is well known characteristic of wing tone sound. The simulated wing-tone pattern compares very well with the recorded and processed data (\figref{fig:sim}d), though the recorded data showed a substantial cycle-to-cycle variation which is about $\pm$ 10 dB. The present results showed that the wing- tone from individual mosquito can be predicted with reasonable accuracy by using the kinematics reconstruction and aerodynamic/acoustic simulations, and this enables one to investigate the wing-tone communications in courtship and mate chasing.


Within the free-flying subclass, wing kinematics are constructed for six individuals (three males and three females). These are then used to set up aerodynamic and aeroacoustic simulation for each specimen. Here we restrict our discussion to one representative male $(M_1)$ (left wing) and one representative female $(F_1)$ (right wing), both of which are performing a slow descent. Instantaneous snapshots of the wing surface along with the corresponding positions of the 11 points tracked in the digitizing exercise are presented in \figref{fig:force}a. The objective is to highlight the quality of the spatio-temporal wing kinematics constructed for setting up the CFD simulations. Several parameters that categorize different aspects of these two individuals are provided in \tabref{tab:parameters_m1_f1s}. While all 6 individuals translate at virtually uniform - albeit different - velocities, it is clear from video recordings that they are all in free flight. For instance, of the three males recorded, one is descending gradually whereas the other two are in a slow ascending flight. The choice of performing the CFD simulations without externally imposed free-stream velocity ($U_\infty$) is justified by the small advance ratio $J=U_\infty/v_{tip}<0.1$ in \tabref{tab:parameters_m1_f1s}. Despite the translating aspect of their flight, the very small levels of $J$ imply that both individuals ($M_1$ and $F_1$) are roughly hovering.

\begin{table}[!hb]
 \caption{A summary of several parameters that categorized the two cases discussed here: $M_1$ and $F_1$.}
 \begin{center}
 \begin{tabular}{|l|l|l|}
  \hline \textbf{Parameter}	& \boldmath${M_1}$	& \boldmath$F_1$\\
  \hline Wingbeat frequency $f_0$ 	&750 Hz   &450 Hz\\
  \hline Reynolds number $Re_s=\frac{L_s^2f_0}{\nu}$	            &310   &410\\
  \hline Reduced frequency $f^*=\frac{L_c f_0}{v_{tip}}$          &0.12   &0.08\\
  \hline Advance ratio $J= U_\infty / v_{tip}$	            &0.07   &0.04\\
  \hline Stroke amplitude ($\Delta\phi$)  &$36^{\circ}$   &$42^{\circ}$\\
  \hline Pitch ($\Delta\gamma$)	            & $109^{\circ}$   &$148^{\circ}$\\
  \hline Mean lift force $F_z$ 	            &3.5 $\mu$N  &9.1 $\mu$N\\
    \hline Mean drag force $F_y$	            &1.1 $\mu$N   &90.0 $\mu$N\\
  \hline Mean lift by upstroke           &54 \%   &48 \%\\
  \hline  
 \end{tabular}
\end{center}
 \label{tab:parameters_m1_f1s}
\end{table}

\begin{figure}[!h]
    \centering
    \begin{subfigure}[b]{0.5\columnwidth}
 \centerline{\framebox{
 \includegraphics[width=0.9\textwidth]{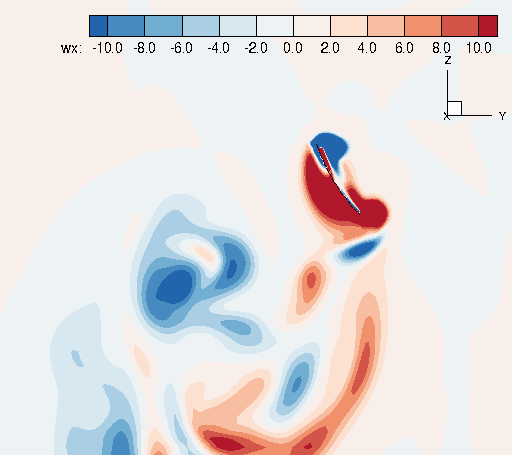}}}
 \caption{$t/T=0.16$}
    \end{subfigure}~
    \begin{subfigure}[b]{0.5\columnwidth}
\centerline{\framebox{
 \includegraphics[width=0.9\textwidth]{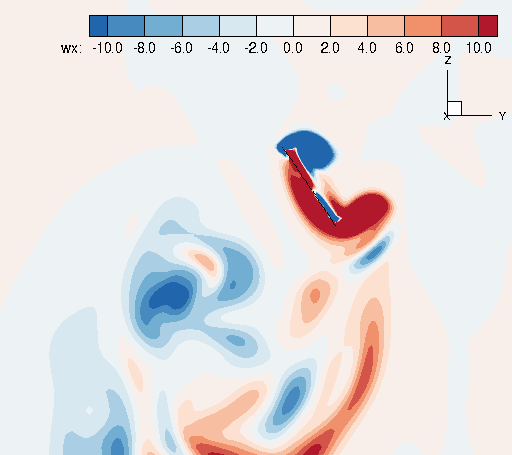}}}
 \caption{$t/T=0.22$}
    \end{subfigure}\\
    \begin{subfigure}[b]{0.5\columnwidth}
 \centerline{\framebox{
 \includegraphics[width=0.9\textwidth]{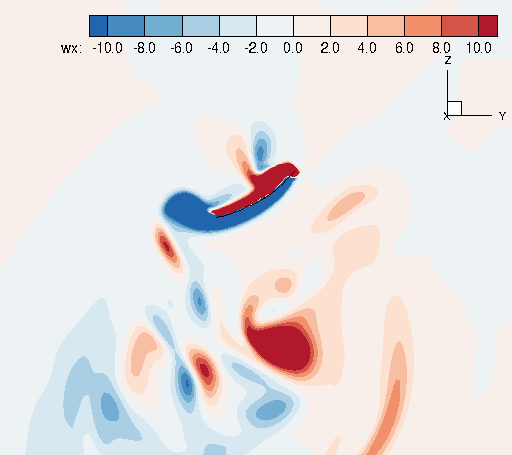}}}
 \caption{$t/T=0.74$}
    \end{subfigure}~
    \begin{subfigure}[b]{0.5\columnwidth}
\centerline{\framebox{
 \includegraphics[width=0.9\textwidth]{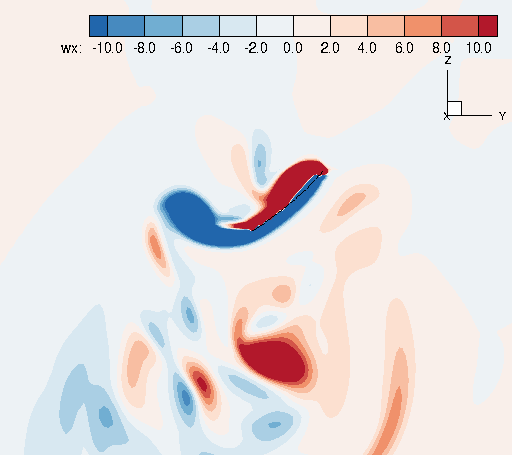}}}
 \caption{$t/T=0.80$}
    \end{subfigure}
    \caption{Spanwise vorticity ($\omega_x$) for case $M_1$ in a $yz-$plane at $x/L_s=0.7$. $(a)$ Peak drag during downstroke, $(b)$ peak lift during downstroke, $(c)$ peak lift during upstroke and $(d)$ peak drag during upstroke.}
    \label{fig:vis_xl0p7}
\end{figure}

The temporal variation of lift and drag forces along the flapping cycle for $M_1$ and $F_1$ is presented in \figref{fig:force}b-c. The higher mean force levels for case $F_1$ are expected as it operates at a higher Reynold number ($Re_s$). The observation from these two individuals that peak lift occurs near the middle of the upstroke is consistent with the result by \cite{BomphreyNakPhiWal17,SeoHedrickMittal19}. But this observation does not extend to all individuals tested. However, what is consistent across all free flying individuals simulated is the fact that mean lift produced during the downstroke ($0 \le t/T \le 0.5$) and upstroke ($0.5 \le t/T \le 1.0$) is essentially identical. Vortical structures that develop around the wing for case $M_1$ in a cross-spanwise plane at $x/L_s=0.7$ and at four key time instants in the flapping cycle are presented in \figref{fig:vis_xl0p7}. It is evident that peak lift during each half stroke coincides with the moment at which the trailing edge vortex (TEV) is around its largest size and closest to the trailing edge. Despite the apparent significance of the TEV, the major contributor to aerodynamic lift for free flying mosquitoes is the thin `underside shear layer (USL)’ that develops on the pressure side of the wing. The USL in turn feeds the TEV. In contrast to the TEV, a dominant leading edge vortex that would grow before shedding is not observed. This is not surprising given the shallow strokes and the rapid pitching component of the overall flapping motion.

\begin{figure}[H]
    \centering
    \begin{subfigure}[b]{1\columnwidth}
 \centerline{\framebox{
 \includegraphics[width=0.7\textwidth]{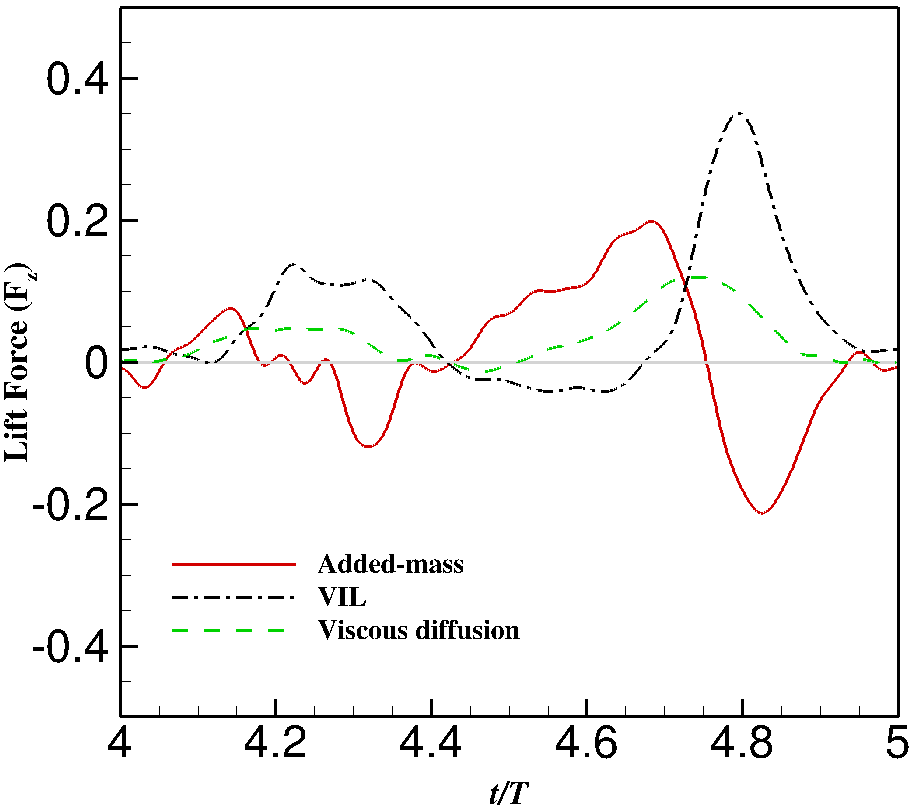}}}
 \caption{}
     \end{subfigure}\\
    \begin{subfigure}[b]{1\columnwidth}
 \centerline{\framebox{
 \includegraphics[width=0.7\textwidth]{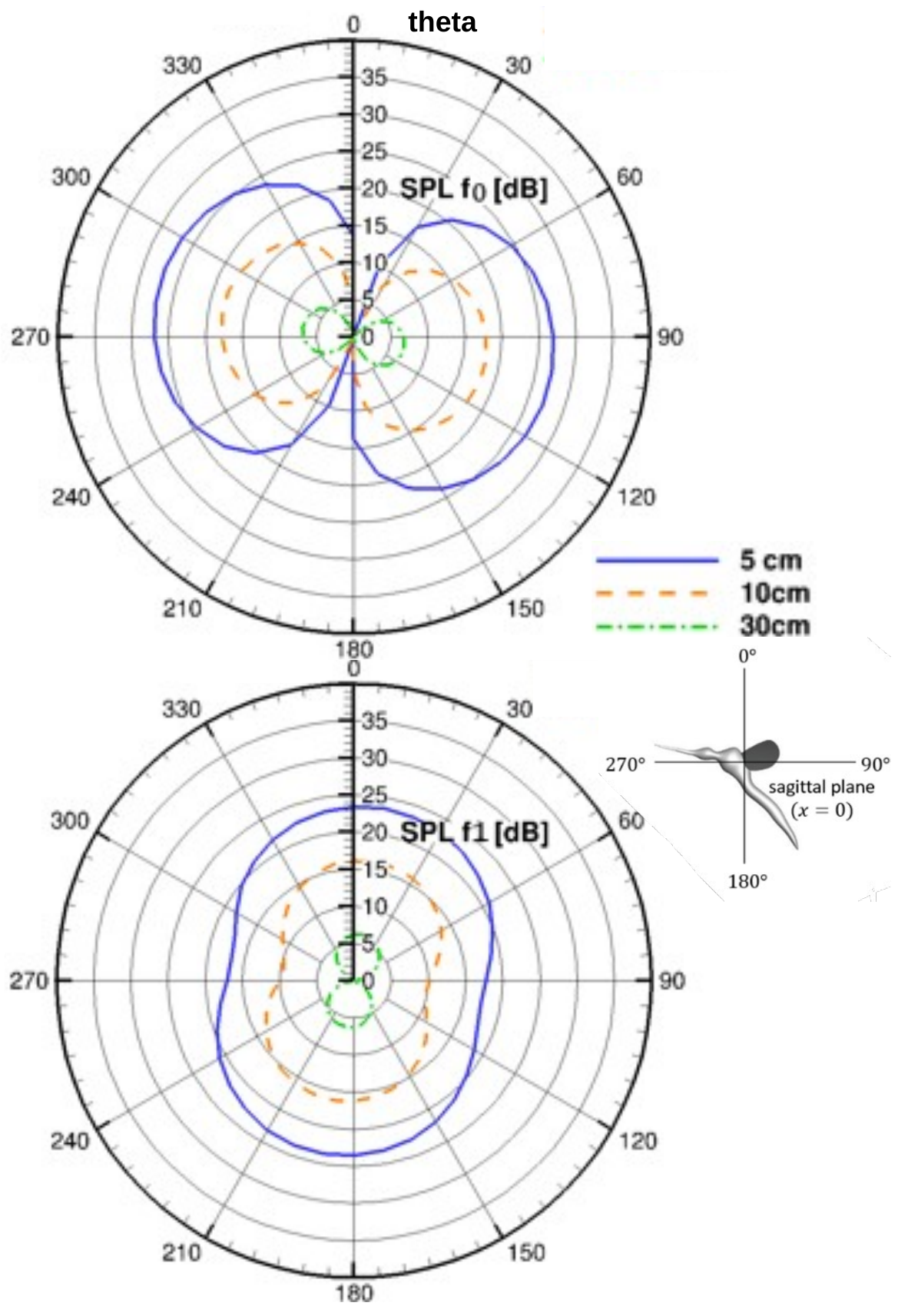}}}
 \caption{}
    \end{subfigure}    
    \caption{$(a)$ Temporal variation of different components of the FPM that constitute the total lift force along a flapping cycle of case $M_1$. Normalization of the ordinate is by length of the wingspan $L_s$ and wingbeat frequency $f_0$. $(b)$ Directivity patterns of the SPL (in dB) at $f_0$ (top) and $f_1$ (bottom) for case $M_1$.}
    \label{fig:fpm_m1}
\end{figure}

The importance of the USL is confirmed by applying the Force Partitioning Method (FPM) \cite{Menon2021, 2022APS, fpm_mostafa} to the resulting flow field. The FPM offers a mathematical partitioning of the total pressure force on the wing surface into components due to vortical regions, added-mass effects, and viscous momentum diffusion. Perhaps more importantly, the FPM allows one to quantitatively identify dynamically important vortical regions within the aerodynamic field. Further details of the FPM are provided by \cite{Zhang2015}. Among the three constituent elements of the mean pressure-induced lift, the term due to the vortical regions, also referred to as the vortex-induced lift (VIL), is the largest contributor at $61\%$ for case $M_1$ (\figref{fig:fpm_m1}a). While the other two components also induce a positive lift in the mean sense, with lift due to viscous momentum diffusion being the dominant one at $32\%$, the contribution by the added-mass term has a negative correlation with the VIL term. It is worth highlighting that the added-mass term is independent of $Re_s$, whereas the contribution by viscous momentum diffusion would decrease progressively with increasing $Re_s$ (i.e. higher wingtip velocity). Furthermore, there is non-negligible cycle-to-cycle variation in the 11 experimentally tracked points sited on the wing. This is accounted for by performing CFD simulations for every individual using two distinct sets of wing kinematics, where each set represents a different cycle of the flapping motion (\figref{fig:force}b-c).

The directivity pattern of the sound pressure level (SPL) at the fundamental $(f_0)$ and superharmonic frequencies in the sagittal plane $(x=0)$, as predicted by the FW-H equation for $M_1$, is presented in \figref{fig:fpm_m1}b. Apart from their symmetric bimodal shape, we see that the SPL radiates mostly in the horizontal and vertical directions at $f_0$  and $f_1$ frequencies, respectively. As noted by \cite{SeoHedrickMittal19}, the former is a consequence of strong temporal fluctuations of the drag force at $f_0$, while the latter is a result of dominant temporal fluctuations of the lift force at $f_1$.

\section{Acknowledgments}
This research was funded by Human Frontier Science Program Research Grant No. RGP0038/2019.
The authors would like to acknowledge Shahida Begum 
for providing mosquito eggs, Natalie Morley and Simon Springate for rearing  mosquitoes and Tyson Hedrick for support on using the DLTdv digitalizing software.

\bibliography{bib-lio}

%
%
%

\end{document}